# Microstructural rearrangements and their rheological implications in a model Thixotropic Elasto-Visco-Plastic (TEVP) fluid

*Safa Jamali[1], Gareth H. McKinley[2] and Robert C. Armstrong[1]*

1- Department of Chemical Engineering, Massachusetts Institute of Technology,
77 Massachusetts Avenue, Cambridge, Massachusetts 02139, USA

2- Department of Mechanical Engineering, Massachusetts Institute of Technology,
77 Massachusetts Avenue, Cambridge, Massachusetts 02139, USA

We identify the sequence of microstructural changes that characterize the evolution of an attractive particulate gel under flow and discuss their implications on macroscopic rheology. Dissipative Particle Dynamics (DPD) is used to monitor shear-driven evolution of a fabric tensor constructed from the ensemble spatial configuration of individual attractive constituents within the gel. By decomposing this tensor into isotropic and non-isotropic components we show that the average coordination number correlates directly with the flow curve of the shear stress vs. shear rate, consistent with theoretical predictions for attractive systems. We show that the evolution in non-isotropic local particle rearrangements are primarily responsible for stress overshoots (strain-hardening) at the inception of steady shear flow and also lead, at larger times and longer scales, to microstructural localization phenomena such as shear banding flow-induced structure formation in the vorticity direction.

Thixotropic Elasto-Visco-Plastic (TEVP) materials are a broad class of structured fluids that include (but are not limited to) most colloidal gels [1], nano emulsions [2], crude oils [3, 4], and many biological systems such as blood clots and actin networks [5, 6]. As a result of their complex underlying microstructure, TEVPs exhibit a wide range of rich and complex thermo-mechanical properties: Below a critical stress, the microstructural network formed by individual particles remains intact and resists large deformations by external forces. At this stage the macroscopic response of the material is similar to that of a viscoelastic solid. By progressively increasing the applied load, the material reaches its "yield stress" and starts to flow [7]. At this point the particle network that is responsible for solid-like response of the macroscopic sample undergoes plastic rearrangements over an increasingly wide range of length scales [8]. Upon complete yielding of this network, plastic flow results ultimately in a viscous-like response; however, as a result of constant formation and breakage events, the particle-level microstructure continues to evolve giving rise to thixotropic behavior. The many-body nature of the problem means that local forces exerted on a single particle change its energy landscape, which consequently defines its subsequent association/dissociation rate to neighboring particles [9, 10]. When combined with multi-body hydrodynamic effects in these fluids [11], the resulting microstructure-flow relationship becomes complex and may show long time-scale transient behavior and multiple steady states [1, 12]. This leads to a wide range of time-dependent responses that can also be observed, including micro-phase separation [13], vorticity aligned structure formation [14-16], local rigid plug formation and shear banding [17], plus shear-induced rejuvenation of the particle network [18].

Although the general form of the flow curve (relating the shear stress to shear rate) and some transient phenomena of TEVP fluids can be described by detailed continuum models [3, 19-21], a clear and fundamental understanding of the underlying mechanisms requires connecting macroscopic material properties to events that happen at the microstructural level. Experimentally, making this connection is very challenging because monitoring of the entire microstructure under flow is essential, together with elaborate protocols to produce well-defined initial conditions [22]. Thus, numerical simulations can play a crucial role in providing a detailed picture of the local structural evolution in these complex fluids.

Bridging the gap between the particle-level interactions and the macroscopic rheological response of a complex fluid requires access to mesoscopic length/time scales spanning the microscale, where the building blocks of a system are described explicitly, while concomitantly accessing much longer time/length scales relevant to macroscopic phenomena. To span this wide range of length/time scales, we perform DPD [23] simulations on model TEVP fluids consisting of 10 *vol%* 2D attractive hexagonal solid particles designed to represent the waxy crystalline particles observed in a waxy crude oil [3, 24]. The DPD formalism inherently preserves multi-body hydrodynamics by conservation of mass and momentum both locally and globally [25, 26]. Furthermore, by incorporating relevant interaction potentials for attractive colloids, this technique can provide a realistic representation of a TEVP fluid.

In DPD, the equation of motion of a particle is written based on pairwise interaction potentials: $m_i \frac{d\mathbf{v}_i}{dt} = \sum_{j=1}^{N} \mathbf{F}_{ij}^D + \mathbf{F}_{ij}^R + \mathbf{F}_{ij}^C$, that act on each individual particle [27]. Where $N$ is the number of interacting neighbors around particle $i$. The dissipative, $\mathbf{F}_{ij}^D$, and random, $\mathbf{F}_{ij}^R$, interactions define the temperature in the system and satisfy the fluctuation-dissipation theorem, whereas the conservative force, $\mathbf{F}_{ij}^C$, is a sole function of the instantaneous spatial configuration of two interacting particles. In traditional DPD [23, 27-29], this force is written as a soft linearly-decaying repulsive potential: $\mathbf{F}_{ij}^C = a_{ij} \omega_{ij}^C(\mathbf{r}_{ij}) \mathbf{e}_{ij}$, where $a_{ij}$ is the strength of the inter-particle repulsion. The attractive nature of local particle-particle interactions in TEVP materials, such as colloidal gels, demands that a modified conservative potential be employed [30]. Thus we modify the interaction potential between the solid particles to a Morse potential: $\mathbf{F}_{ij}^{C} = \Gamma \kappa (e^{-\kappa r_{ij}} - e^{-2\kappa r_{ij}}).\mathbf{e}_{ij}$, where $\Gamma$ is the depth of the attraction well and $\kappa^{-1}$ is the characteristic range of attraction [30]. Because of the short-range (0.2 of the particle size) nature of the attractive potential, $\Gamma$ provides a direct measure of the attraction strength between the particles. We have performed simulations with the attraction strength of $3k_B T \leq \Gamma \leq 50 k_B T$, representing a range of weak to strong interactions between the particles with $\kappa^{-1} = 0.15$. For details of the simulation technique, interactions, parameters, non-dimensional scaling, and imposed flow conditions please refer to the SI.

Colloidal networks obtained from attractive interactions between the constituents, are subject to simple shear flows, where the constant velocity of the upper and lower boundary cells, $v_x = \dot{\gamma} y$, induces a linear velocity profile within the calculation box for a simple fluid with $\dot{\gamma} = 2V_{Wall} / L_y$. We evaluate the steady state value of the corresponding shear stress, $\sigma_{xy}$, at long times; and Fig.1 shows the general flow curve for an attractive gel ($\phi = 0.1$ and $\Gamma = 20 k_B T$) as well as the initial evolution of shear stress as a function of accumulated strain deformation. The first noticeable feature of Fig.1a is that the key experimentally-observed feature [3, 4] corresponding to yielding behavior in a TEVP gel is reproduced: the shear stress approaches a finite plateau at low dimensionless values of the imposed shear rate, $\dot{\gamma} \leq 0.1$. This is in agreement with theoretical predictions of Snabre and Mills [20] for attractive colloidal systems under steady shear flow. Details of the model and the values of the parameters obtained from fitting the theory to our simulation are presented in the SI. Fig.1b shows the time-dependent flow behavior of the gels at the startup of steady shear, where the shear stress is plotted as a function of imposed strain for different shear rates. A marked stress overshoot is evident regardless of the imposed shear rate at a dimensionless strain of $\gamma \simeq 0.2$. This is consistent with several experimental measurements [13, 31, 32] and computational predictions of these materials at flow start up [8, 33]. The origin of this stress-overshoot is commonly associated with rearrangement of particles within the fluid in conjunction with stretching of the microstructure along the direction of maximum extension [8]. Numerical simulations of viscoelastic constitutive equations have also shown that stress overshoots are commonly associated with onset of hydrodynamic instabilities and the development of inhomogeneous shearing profiles in shear complex fluids [17, 34, 35]. In Fig.1c we show space-time diagrams of the velocity profiles, $v_x(y,t)$, across the simulation cell together with final profiles at $t$=250 presented in Fig.1d for a range of shear rates. The profiles in Fig.2d are clearly not linear across the box and show strong evidence of flow inhomogeneity at intermediate deformation rates, $0.1 < \dot{\gamma} < 1.0$. At smaller deformation rates, $\dot{\gamma} < 0.1$, significant slip is observed with regards to velocity of the moving boundaries, whereas at the highest rates of $\dot{\gamma} > 1$ no evidence of inhomogeneity is observed across the calculation cell. We seek to understand how evolution of the material on the microscale gives rise to these observed kinematic inhomogeneities.

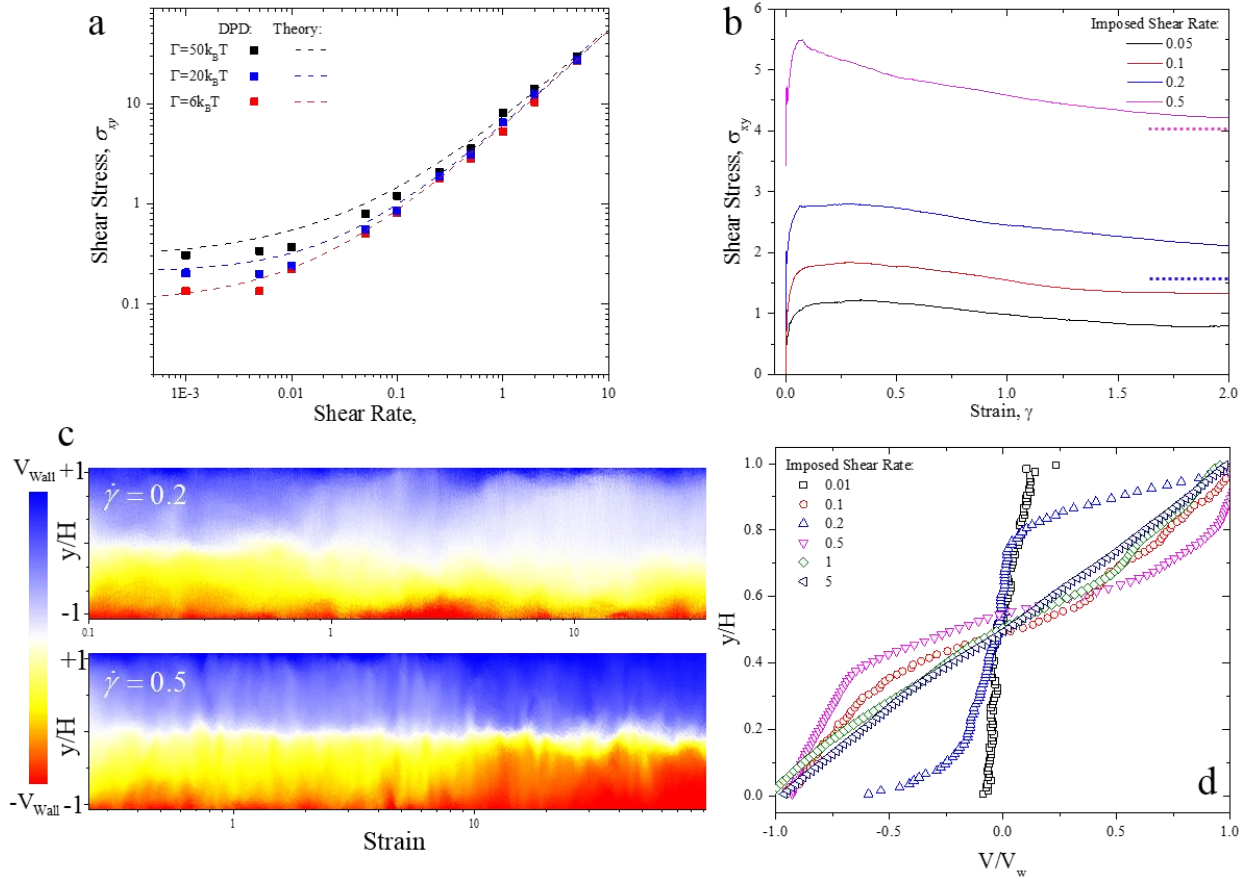

*Figure 1. a) Shear stress versus shear rate, compared to [20], b) Stress overshoot response in flow start up simulations for different shear rates, dashed-lines show the steady state asymptotes at sufficiently large deformations, c) Space-time diagrams of the velocity field $v_x(y,t)$ in the gel under two different strain rates showing flow instability and spatial inhomogeneities at large strains/times, and d) Final averaged velocity profiles across the calculation box, $\bar{v}_x(y)$, for a range of shear rates.*

To do so, we characterize the microstructure of the gel network through the fabric tensor, $\mathbf{Z}$ [36-38]. As two interacting particles approach within the attraction range, they form particle bonds. To monitor the collective dynamics of the system we define, $\mathbf{Z}$, from the local fabric [36-38] of single particles ($\mathbf{Z}^p = \sum_i \mathbf{n}_i \otimes \mathbf{n}_i$), where $\mathbf{n}_i$ is the unit vector of the center-center line between a particle and its $i$-th bond forming neighbor. Averaging $\mathbf{Z}^p$ over the particle ensemble will yield a system-sized fabric tensor: $\mathbf{Z} = \frac{1}{N}\sum_{p=1}^{N} \mathbf{Z}^p$. The trace of this ensemble fabric tensor provides the average number of bonds in a system, $tr\mathbf{Z} \equiv <Z>$, i.e. the coordination number, or the isotropic scalar defining the configuration of each particle and its nearest neighbors [22, 33, 39]. The deviatoric components of this fabric tensor can provide significant information about evolution in the anisotropy of the structure under different conditions. We define the deviatoric measure as: $\boldsymbol{\zeta} = \mathbf{Z} - \frac{tr\mathbf{Z}}{3}\mathbf{I}$ where $\mathbf{I}$ is the identity tensor. The temporal evolution of the deviatoric components of the fabric tensor are presented in Fig.2, as a function of the accumulated strain for different dimensionless shear rates. In our notation, $x$, $y$ and $z$ components correspond to the flow, the velocity gradient and the vorticity directions respectively. The microstructural insights from the evolution of the fabric tensor in Fig.2 can be separated into 3 distinct regimes: (I) Upon flow start up, and within the linear viscoelastic regime (at small strains of $\gamma \leq 0.1$), the particulate network remains rather isotropic with little change in the microstructure and the corresponding stress grows linearly with strain (Fig.1b). (II) At $\gamma \simeq 0.1$ the material response deviates from linear viscoelasticity, and the increase in the $\zeta_{xx}$ component suggests that the inter-particle bonds start to rearrange preferentially into the $x$-direction. At these deformation levels of $0.1 \leq \gamma \leq 10$, bonds are also increasingly broken in the $y$ direction, as evident from the decrease in $\zeta_{yy}$. (III) At larger strains of $\gamma \geq 10$, persistent transient structures are observed, with both $\zeta_{xx}$ and $\zeta_{yy}$ showing steady plateaus (after passing through a maximum and a minimum respectively); and in addition $\zeta_{zz}$ starts to increase, indicating the slow evolution of structures aligned in the vorticity direction.

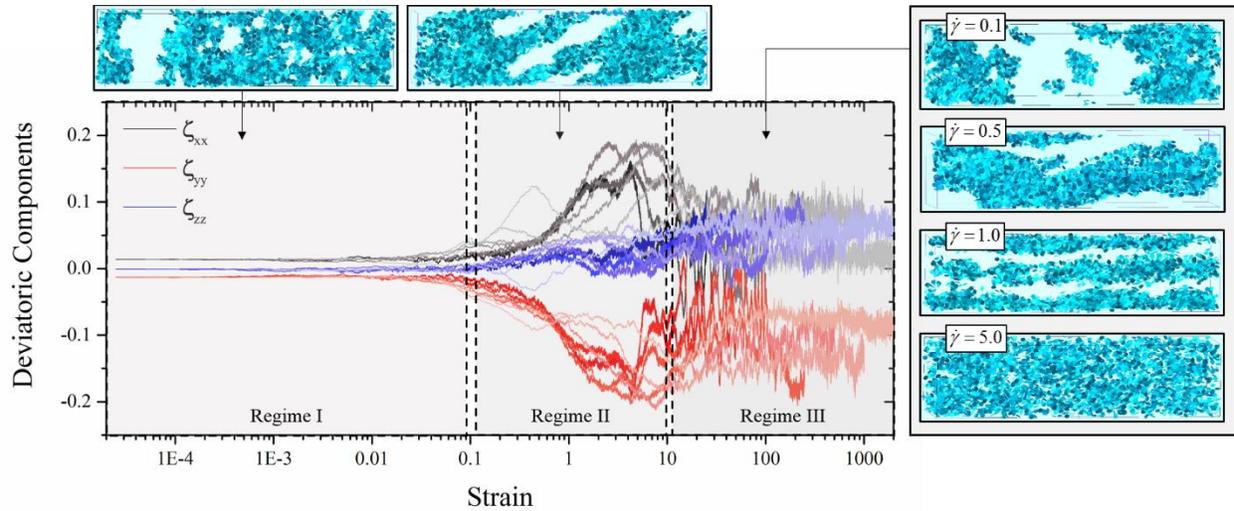

*Figure 2. Deviatoric components of the fabric tensor as a function of accumulated strain for different shear rates (color increments from dark to light indicate increase in shear rate), accompanied with snapshots of representative microstructures in different strain regimes.*

It should be noted that although the general evolution in the microstructural anisotropies are similar regardless of imposed strain rates, the morphology of the particulate network on larger length scales varies significantly depending on the rate of shearing. This is evident in the final images shown for $\gamma \simeq 100$ in Fig.2. At high shear rates ($\dot{\gamma} = 5$) the microstructure becomes homogeneous, whereas at lower deformation rates of 0.1, 0.5 and 1, small clusters, plug-like phase separated morphologies and string-like layers of particles are observed respectively.

Fig.3 shows the evolution in the off-diagonal shear component of the fabric tensor, $Z_{xy}$, as well as the coordination number, $tr\mathbf{Z}=<Z>$, as functions of shear rate and accumulated strain. As evident in Fig.3a the shear component of the fabric tensor (fig. 3), follows the shear stress regardless of the applied deformation rate, and undergoes an overshoot at $\gamma \simeq 1$. As expected, this confirms that the deviation in linear viscoelastic response of a TEVP fluid originates from the re-arrangements occurring within its microstructure. In addition, the other non-diagonal components show similar universal changes for the same range of strains (see Fig.S8 in the Supplementary Information).

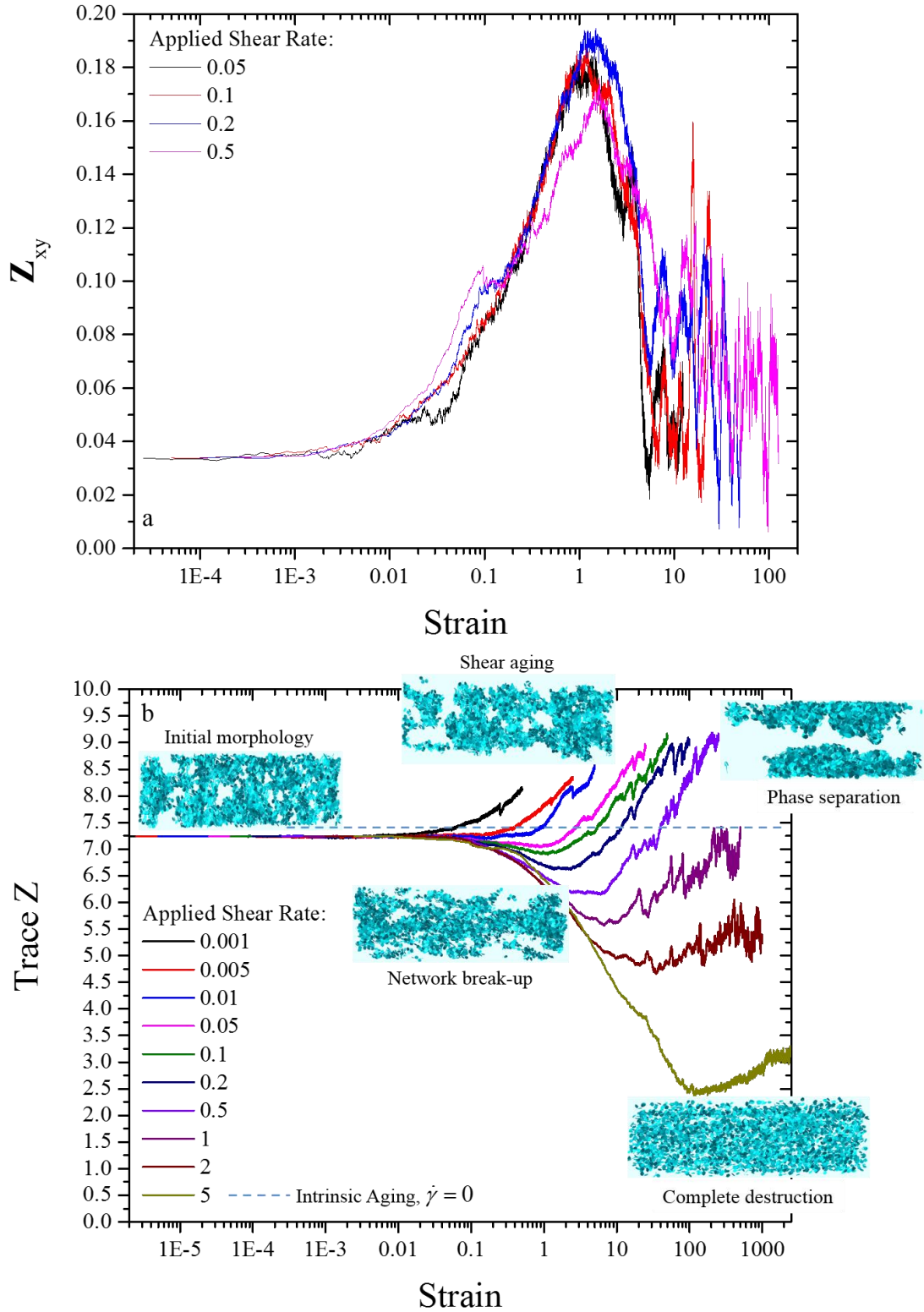

*Figure 3. a) Overshoot in shear component of the fabric tensor in flow start up simulations, and b) Trace of the fabric tensor (coordination number) for different shear rates. Inserts show representative microstructures in different flow regimes (compaction, shear rejuvenation, network break-up and etc.).*

However, following the trace of the fabric tensor, $tr\mathbf{Z}$, for the same range of shear rates allows us to clearly differentiate between different material responses (Fig.3b). At low strain rates, prolonged shearing results in a net formation of new bonds and compaction of the structure; and thus this regime is most commonly referred to as "shear-aging" [30], and $tr\mathbf{Z}$ increases monotonically. By increasing the shear rate above the critical strain rate for macroscopic yielding, the shear stresses initially break the network into smaller clusters and individual particles, resulting in a steady decrease in the coordination number. However, new intra-cluster bonds form on longer time scales; and this results in larger coordination numbers than the initial configuration as these secondary structures begin to form. This regime is most commonly referred to as shear-rejuvenation [40], as the initial structure is progressively destroyed and a new one is formed. For high accumulated deformations, large secondary structures are formed, which are highly anisotropic (as confirmed by the individual diagonal components of the fabric tensor presented in Fig.2 and the supplementary materials, see figure S8). At the highest shear rates, $\dot{\gamma} \geq 5$, the macroscopic response corresponds to steady viscoplastic flow (Fig.1a) and the particulate network is effectively broken into small homogenous structures with very low coordination numbers.

Heterogeneities in the microstructure of the fluid under flow result in inhomogeneous fabric tensor measurements at the particle level. Thus, in addition to presenting ensemble-averaged measures of the fabric components as shown in fig.2 and fig.3, we also show in figure 4 the distribution of coordination numbers in the fluid compared to this distribution prior to inception of flow.

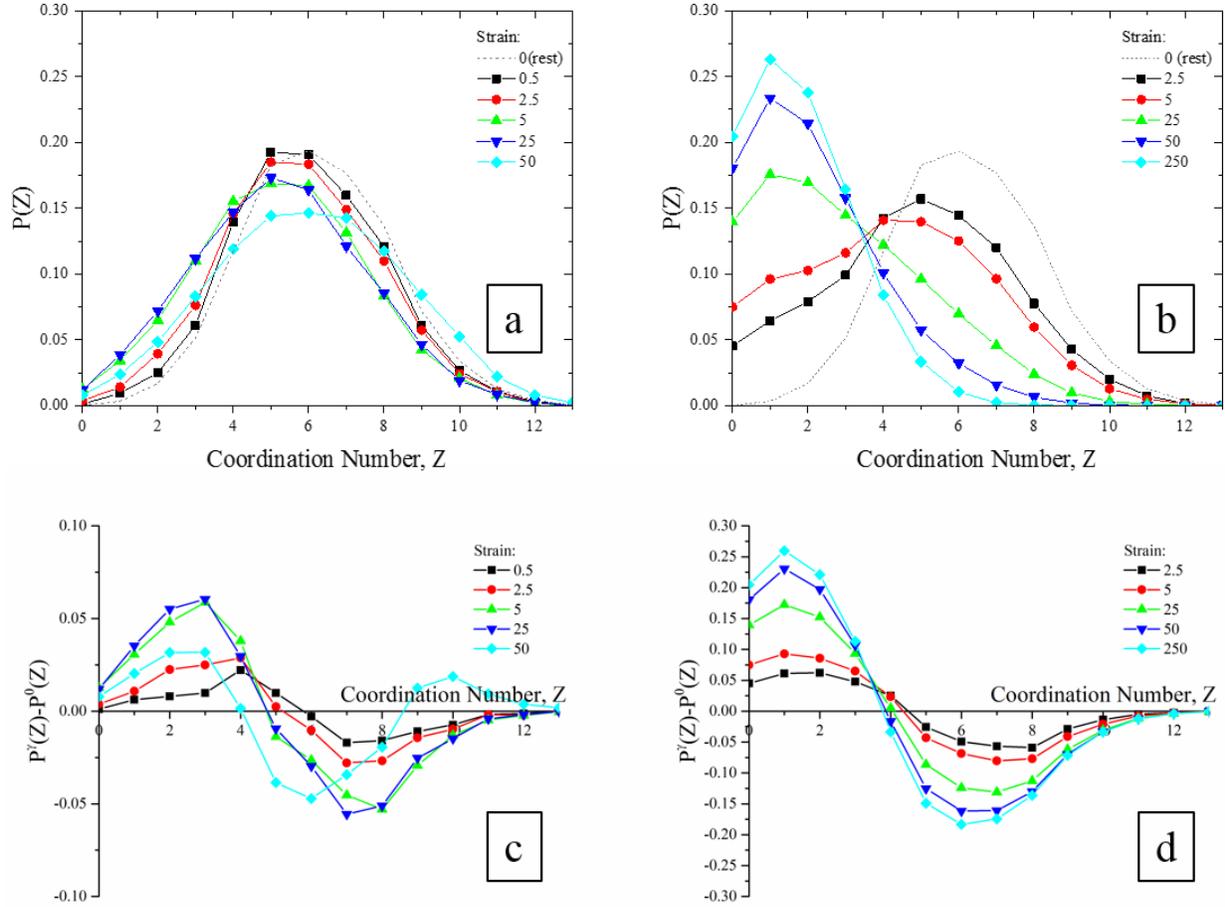

*Figure 4. Evolution in the distribution of coordination number for a network of 4500 attractive particles ( $\phi = 0.1$ and $\Gamma = 20k_BT$ ) at different strains under shear rates of 0.5 (a) and 5.0 (b). Change in the coordination number relative to the quiescent distribution for shear rate of 0.5 (c) and 5.0 (d).*

Examining the distribution of coordination number for the gel that undergo break-up and reformation of transient structures at intermediate shear rate of 0.5 in fig. 4(a,c) shows that at short times ($\gamma < 10$) population of particles with $Z \geq 6$ progressively decreases in the system and number of particles with $Z \leq 5$ increases. However, at longer times when secondary structures are formed, the coordination number distribution relative to quiescent condition, $P^\gamma(Z) - P^0(Z)$, shows two maxima and a minimum, indicating that there are local flocculated sub-domains, coexistent with regions in which the particles have smaller coordination numbers. However, by increasing the shear rate to 5.0, a gradual destruction of the particle network is observed, in agreement with results presented in figure 3, and with the experimental measurements by Hsiao *et al.* [22].

In summary, we have shown that the particle networks reproduced by DPD for soft attractive particles capture the key rheological features of a model TEVP fluid and are consistent with the predicted form of phenomenological models for attractive systems. Calculating a fabric tensor from the ensemble of individual particle-particle bonds within a gel-forming colloidal system provides additional insight about the microstructural changes that occur under flow. Particle re-arrangements at moderate strains, $0.1 \leq \gamma \leq 2$, lead to an overshoot in the shear stress and the $Z_{xy}$ component of the fabric tensor also undergoes a similar overshoot. To decouple changes in the average isotropic and anisotropic contributions to the microstructure of the attractive gel under shear we calculate the deviatoric fabric tensor, $\zeta$. This shows evidence of bond re-arrangements in the flow direction and the velocity gradient direction, resulting in a net positive value of $\zeta_{xx} - \zeta_{yy}$. At higher strain units, $\gamma \geq 10$, our three-dimensional calculations of $\zeta_{zz}$ also show development of secondary structure formation in the vorticity direction. This coincides with shear-rejuvenation and a decrease in the average coordination number of the microstructure (Fig.3b) and in the development of the strongly inhomogeneous shearing velocity profiles we have documented in the space-time diagrams (Fig.1c,d).

Our results demonstrate that the DPD method can provide a systematic route to bridging the gap between direct observations [22] of microstructure and measurements of the macroscopic rheological response of a TEVP fluid to an imposed deformation rate. Our steady-state simulations show quantitative agreement with closed-form constitutive equations for yield stress fluids, but such models provide no insight into transient responses of such materials. Recent microscopic measurements, as well as large scale numerical simulations, have begun to focus on the start-up of steady shear flow and have clearly documented the evolution of particulate level microstructure under deformation [8, 18, 22, 33, 39, 41, 42]. The wide range of time and length scales accessible by DPD simulations, coupled with inclusion of hydrodynamic interactions between particles [25, 43], now enables systematic exploration of the evolution of microstructural and rheological properties for TEVP materials as well as validation of empirical constitutive equations proposed for such systems [21, 44-47].